\documentclass[preprintnumbers,amsmath,amssymb,usenatbib,nofootinbib,superscriptaddress,showkeys,showpacs,altaffilletter,11pt]{revtex4-1}
\usepackage{amsmath}
\usepackage{amsfonts,color}
\usepackage{amssymb,float}
\usepackage{graphicx}
\usepackage{appendix}
\usepackage[colorlinks]{hyperref}
\usepackage{xcolor}
\usepackage{orcidlink}
\usepackage{epsf}
\usepackage{bm}
\usepackage{epstopdf}
\usepackage{natbib}
\usepackage{lipsum}
\usepackage{autobreak}

\usepackage[colorlinks]{hyperref}

\begin{document}

\title{Extended Bose-Einstein condensate dark matter in viscous Gauss-Bonnet gravity}

\author{E. Mahichi}
\email{e.mahichi@iauamol.ac.ir}

\author{Alireza Amani\orcidlink{0000-0002-1296-614X}}
\email[Corresponding author:~]{a.r.amani@iauamol.ac.ir}

\author{M. A. Ramzanpour}
\email{m.ramzanpour@iauamol.ac.ir}
\affiliation{Department of Physics, Ayatollah Amoli Branch, Islamic Azad University, Amol, Iran}

\date{\today}

\begin{abstract}
In this paper, we study the $F(R, G)$ gravity model with an interacting model by flat-FRW metric in a viscous fluid. We consider that the universe dominates with components of dark matter and dark energy. This means that the dark matter component derives from Extended Bose-Einstein Condensate (EBEC) and the components of dark energy arise from the $F(R, G)$ gravity. After obtaining the Einstein equation, the energy density and the pressure of dark energy are written in terms of the geometries of the curvature and the Gauss-Bonnet terms, and components of dark matter and viscous fluid. Also, the corresponding continuity equations are written with the presence of interaction terms. In what follows, we employ the EBEC regime instead of the normal dark matter by the dark matter Equation of State (EoS) as $p_{dm} = \alpha \rho_{dm} + \beta \rho_{dm}^2$, which arises from the gravitational form. The EoS can be expressed from the perspective of the virial expansion, in which the first and second terms represent normal dark matter and quantum ground state. Next, the corresponding Friedmann equations reconstruct in terms of the redshift parameter, then by using the scenario of the power-law cosmology for the scale factor, we fit the present model with the Hubble amounts of 51 supernova data by the likelihood analysis. In that case, we acquire the cosmological parameters of dark energy in terms of the redshift parameter, and by plotting these graphs, we see that the universe is currently undergoing an accelerated expansion phase. Finally, we investigate the stability of the present model with the sound speed parameter.
\end{abstract}



\maketitle
\section{Introduction}\label{I}
At the beginning of the last century, various attempts were made to explore beyond our galaxy, which led to the discovery of the expansion of the universe that Friedmann and Hubble had a great contribution in this discovery. Friedmann described the theory from the Einstein field equation with his famous equations called Friedmann equations in the Friedmann–Lema\^{\i}tre–Robertson–Walker (FLRW) metric and a perfect fluid. This expansion was confirmed by Hubble-Lema\^{\i}tre law, which states that the universe is expanding at a constant speed and is proportional to the distance of the galaxies from Earth. Since the velocity of the galaxies is determined by their redshift, so they moving away with acceleration from the Earth. This means that the universe is undergoing an accelerated expansion, that this issue was discovered by type-Ia supernova, cosmic microwave background, and large scale structure \cite{Riess_1998, Perlmutter_1999, Bennett_2003, Tegmark_2004}. This discovery shows that the expansion velocity of the universe is under a mysterious and unknown force
and is introduced as a hypothetical energy called dark energy which is necessary to have a strong negative pressure. Therefore, a number of scenarios studied for description of dark energy or same the late time acceleration entitled the cosmological constant \cite{Weinberg-1989, Ng-1992}, scalar fields (quintessence, phantom, quintom, tachyon, and etc) \cite{Chiba-2000, Kamenshchik-2001, Caldwell-2002, Singh-2003,Guo-2005, Wei-2005, Setare1-2009, Amani-2011, Sen-2002, Bagla-2003, Sadeghi1-2009, Setare-2009, Amani-2013, Amani-2014, Battye-2016, Li-2012}, modified gravity \cite{Dolgov-2003, Faraoni-2006, Nojiri-2007, Iorio-2016}, holography and agegraphics \cite{Li-2004, Wei-2009, Amani1-2011, Campo-2011, Hu-2015, Wei-2007, Jamil-2010, Jawad-2013}, bouncing theory \cite{Shtanov-2003, Sadeghi-2009, Sadeghi-2010, Amani-2016, Singh-2016}, teleparallel gravity \cite{Capozziello-2011, Myrzakulov-2011, Pourbagher_2019, Rezaei-2017, pourbagher1-2020} and braneworld models \cite{Sahni-2003, Setare-2008, Brito-2015}.

Since the nature of dark energy is still unknown, various studies are underway to describe it, which is the motivation for the present study. Nowadays, modified gravity is an appropriate alternative instead of the standard gravity model as a source of dark energy. For this purpose, we choose a combination of $f(R)$ and $f(G)$ gravity entitled $f(R, G)$ gravity which can be an appropriate candidate to describe the universe evolution from early time to late time \cite{Nojiri1-2005, Nojiri-2005, Guo-2009, Bamba-2010, Garcia-2011, Dombriz-2012, Atazadeh-2014, Makarenko-2013, Kofinas-2014, Laurentis-2015, Bamba-2017, Carloni-2017, Shamir-2017, Glavan-2020, Elizalde-2020, Jimenez-2020, Odintsov-2020}. Note that $f(R)$ gravity is an arbitrary function of Ricci scalar, $R$, and $f(G)$ gravity is a general function of Gauss-Bonnet term, $G$, which $G$ arises from string theory prediction. Thus, $f(R, G)$ gravity able to adapt to recent observational data for the accelerated universe and to transition from deceleration to acceleration \cite{Chern-1944, Baojiu-2007}.

As we know, the universe consists of visible parts and dark parts, in which visible parts consist of every visible object in the universe, and the dark parts consist of dark energy and dark matter. Therefore, one of the dark and mysterious parts of the universe is dark matter that is not visible in the spectrum of electromagnetic radiation. The gravitational lensing and the cosmic microwave background \cite{Kaiser-1993, Massey-2010, Seljak-1999, Arkani-2009} are the observational evidence for existence of dark matter which supposed to describe the difference between the calculated mass for giant celestial bodies by the two ways of gravity and the luminous matter in them such as gas, stars, and dust. This means that experimental evidence suggests that dark matter was emerged by gravitational pull on normal matter. Thus the gravitational effect between them describes the formation and evolution of galaxies and clusters and large-scale structures within the universe. In order to understand the concept of dark matter, we inspire by the Bose-Einstein Condensate (BEC) regime which can be to access an important achievement in understanding the content of the universe . Hence, BEC is a kind of matter state in which a dilute Bose gas is cooled into very low temperature. Because of the low temperature, phase transition happens and a large part of bosons occupy minimum quantum state, and the macroscopic quantum phenomenon appears at that point. Cooled bosons collapse to each other and super particles which have microwave behavior emerge \cite{Anderson-1995, Davis-1995, Bradley-1995}. It means dark matter supposed to be a bosonic gas under the critical temperature that constitute BEC. In some paper, they expressed some connection between BEC and cosmic evolution of dark matter \cite{Harko1-2011, Das-2015, Fukuyama-2008, Li-2014, Boehmer-2007, Das-2018, Suarez-2014, Chavanis-2012, Harko-2011, Dev-2017, Velten-2012, Madarassy-2015, Kain-2010, Harko-2012, Harko-2019, Bettoni-2014, Zhang-2018, Harko-2015, Chavanis-2017, Harko1-2015, Craciun-2020, Castellanos-2020, HajiSadeghi-2019, Atazadeh-2016, Mahichi-2021}. Therefore, it is expected that BEC model can be one of the ways to recognize and describe dark matter.

According to the above, from the perspective of the kinetic theory and the hydrodynamic system, we can respectively obtain EoS of dark matter proportional to first-order and second-order for the energy density of dark matter in which the terms represent normal dark matter and quantum ground state. We note that normal dark matter and quantum ground state come from one-body interaction and two-body interaction between bosonic particles. The corresponding coefficients depend to the average squared velocity, the scattering length, and mass of dark matter (the more details are seen in Sec. \ref{III}). As we know, the EoS of dark matter plays a key role in understanding the universe, for this purpose, the virial expansion can be one of the best ways of writing the EOS of a fluid. In that case, the coefficients of the first and second order for the energy density of dark matter can be taken by values of zero and non-zero, which represent cases of cold dark matter, normal dark matter, and dark matter halo (see Sec. \ref{III} for the more details). It should be note that cold dark matter, normal dark matter, and dark matter halo express $p_{dm} = 0$, $p_{dm} = \alpha \rho_{dm}$, and $p_{dm} = \beta \rho_{dm}^2$ in which $\rho_{dm}$ and $p_{dm}$ are the density energy and the pressure of dark matter, respectively. These issues motivate us to represent Extended Bose-Einstein Condensate (EBEC) model that is a combination model for normal dark matter and quantum ground state. The advantage of this method is that in order to understand one of the components of the universe, dark matter, it simultaneously includes the contribution of one-body interaction and two-body interaction.

For that the universe is considered more realistic, we take the bulk viscosity fluid instead of the perfect fluid. In this case, the bulk viscosity creates internal friction that converts the kinetic energy of the particles into heat \cite{Zimdahl-1996, Naji-2014, JSadeghi-2013, Rezaei-2020}. This in turn can be an appropriate idea to describe the accelerated universe in late time. Therefore, as mentioned above, in this job we consider the source of dark energy and dark matter from viscous $f(R, G)$ gravity and the concept of EBEC, respectively. This means that viscous $f(R, G)$ gravity describes the late universe and EBEC helps to understand the origin of the early universe. To this end, most of our focus is on the late time, so EBEC can help us understand the whole evolution of the universe.

The outline of the current job is organized as the following:\\
In Sec. \ref{II}, we review the general form of the Einstein equation with $f(R, G)$ gravity in the flat-FLRW metric. In Sec. \ref{III}, we consider the nature of dark matter as an EBEC. In Sec. \ref{IV}, we reconstruct the corresponding Friedmann equations in terms of the redshift parameter and also analyze it with observational Hubble data. In Sec. \ref{V}, we consider a specific form of $f(R,G)$ to solve the model, and then, we plot the cosmological parameters in terms of redshift parameter, and Also, we explore the stability analysis. Finally, in Sec. \ref{VI}, we provide a summary of the current model.


\section{Viscous GAUSS-BONNET gravity}\label{II}

In this section, we intend to compose the late time universe by the theoretical framework of a general $f(R,G)$ gravity. In that case, we start by action with an arbitrary function of the gravity and the Gauss-Bonnet gravity in the following form
\begin{equation}\label{action}
S = \int {d^4x \sqrt{-g} \left(\frac{f(R,G)}{2 \kappa^2} + \mathcal{L}_m\right)},
\end{equation}
where $\kappa^2 = 8 \pi G_N$, $G_N$ is the Newton constant, $R$ and $G = R^2 - 4 R_{\alpha \beta} R^{\alpha \beta} + R_{\alpha \beta \gamma \delta} R^{\alpha \beta \gamma \delta}$ are respectively the Ricci scalar and the Gauss-Bonnet term in which $R_{\alpha \beta}$ and $R_{\alpha \beta \gamma \delta} $ being the Ricci and Riemann tensors, and $\mathcal{L}_{M}$ is the matter Lagrangian density. So, we obtain the Einstein equation by taking variation of the action \eqref{action} with respect to the metric as follows:
\begin{eqnarray}\label{action1}
\begin{split}
0 &= \kappa^2 T_{\mu \nu} + \nabla_\mu \nabla_\nu \,\partial_R f - g_{\mu \nu}\, \Box\, \partial_R f + 2 R\, \nabla_\mu \nabla_\nu \, \partial_G f - 2 g_{\mu \nu}\, R\, \Box \,\partial_G f - 4 R_\mu^{~\lambda} \,\nabla_\lambda \nabla_\nu\, \partial_G f \\
& - 4 R_\nu^{~\lambda}\, \nabla_\lambda \nabla_\mu\, \partial_G f + 4 R_{\mu \nu} \,\Box\, \partial_G f + 4 g_{\mu \nu} \,R^{\alpha \beta}\, \nabla_\alpha \nabla_\beta\, \partial_G f + 4 R_{\mu \alpha \beta \nu} \nabla^\alpha \nabla^\beta\,  \partial_G f \\
& - \tfrac{1}{2}\,g_{\mu \nu} \bigl(R\, \partial_R f + G\, \partial_G f - f \bigr) - \bigl(R_{\mu \nu} - \tfrac{1}{2} g_{\mu \nu}\,R\bigr)\, \partial_R f.
\end{split}
\end{eqnarray}
where $\nabla_\mu$, $\Box = \nabla^\mu \nabla_\mu$, and $T_{\mu \nu}$ are the covariant derivative operator, the covariant d’Alembertian operator, and the matter energy-momentum tensor, respectively. In the present paper, we consider the flat-FLRW metric in the following form
\begin{equation}\label{ds21}
ds^2 = -dt^2 + a^2(t) \left(dx^2 + dy^2 + dz^2\right),
\end{equation}
where $a(t)$ is the scale factor.

Now, we consider that the universe dominates by a more realistic fluid instead of a perfect fluid, so that the realistic fluid effects on the evolution of the universe called viscous fluid or bulk viscosity.
Therefore, the effects of cosmic bulk viscosity cause a resistance of fluid flow within the universe that has a direct effect on the cosmic pressure.
In that case, the energy-momentum tensor is
\begin{equation}\label{tmunu1}
T_{\mu \nu} = (\rho_{eff} + p_{eff} + p_{b}) u_\mu u_\nu - \left(p_{eff} + p_{b}\right)\, g_{\mu \nu},
\end{equation}
where $\rho_{eff}$ and $p_{eff}$  are respectively the effective energy density and the effective pressure inside the universe. The pressure of the bulk viscosity and the velocity 4-vectors are introduced as $p_{b} = -3 \xi H$ and $u_\mu$ in which $\xi$ is a positive constant and $u^\mu u_\nu = -1$. Thus, the non-zero components of energy-momentum tensor are found as
\begin{equation}\label{enmom1}
T^0_0 = \rho_{eff}, \,\,\,\,\, T^i_i = - p_{eff} + 3 \xi H,
\end{equation}
where $H = \dot{a}/a$ is the Hubble's parameter in which the index of dot expresses the derivative with respect to cosmic time. By substituting the above equations into Eq. \eqref{action1}, the Friedmann equations obtain in the form
\begin{subequations}\label{fried1}
\begin{align}
\kappa^2 \rho_{eff}& = 3 H^2 \partial_R f -\frac{1}{2} R \partial_R f - \frac{1}{2} G \partial_G f + \frac{1}{2} f + 3 H \partial_R \dot{f} + 12 H^3 \partial_G \dot{f},\label{fried1-1}\\
-\kappa^2 (p_{eff} - 3 \xi H) &= -\left(3 H^2 + \dot{H}\right) \partial_R f + \partial_R \ddot{f} + 2 H \partial_R \dot{f} + \frac{1}{2} f + 8 H \left(H^2 + \dot{H}\right) \partial_G \dot{f} \notag \\
&+ 4 H^2 \partial_G \ddot{f} - \frac{1}{2} G \partial_G f ,\label{fried1-2}
\end{align}
\end{subequations}
where dot indicates the derivative with respect to cosmic time, and also we calculate the terms of curvature, $R$, and Gauss-Bonnet, $G$, by FLRW metric as
\begin{subequations}\label{RG1}
\begin{eqnarray}
& R = 6 \left(2 H^2 + \dot{H} \right),\label{RG1-1}\\
& G = 24 H^2 \left(H^2 + \dot{H} \right).\label{RG1-2}
\end{eqnarray}
\end{subequations}

Using the obtained Friedmann equations, the effective continuity equation is written with the existence of bulk viscosity in the following form
\begin{equation}\label{contineq1}
{\dot \rho _{eff}} + 3 H \left({\rho _{eff}} + {p}_{eff} - 3 \xi H\right) = 0.
\end{equation}

Now, we consider that the universe dominates with components such as dark matter and dark energy. For this purpose, the effective energy density and the effective pressure are expressed in terms of the universe components as
\begin{subequations}\label{roheff1}
\begin{align}
\rho_{eff} &= \rho_{dm} + \rho_{de},\label{roheff1-1}\\
p_{eff} &= p_{dm} + p_{de},\label{roheff1-2}
\end{align}
\end{subequations}
where indices of $dm$ and $de$ demonstrate the dark matter and the dark energy, respectively. Therefore, we write down the continuity equations for the universe components separately as
\begin{subequations}
\begin{eqnarray}\label{contineq2}
&\dot{\rho}_{dm}+3 H (\rho_{dm} + p_{dm}) = Q,\label{contineq2-1}\\
  &\dot{\rho}_{de}+3 H (\rho_{de} + p_{de} - 3 \xi H) = - Q,\label{contineq2-2}
\end{eqnarray}
\end{subequations}
where $Q$ is introduced as an interaction term. We note that the interaction term appears when the energy flow is transferred between the dark components of the universe. It is obvious that the quantity $Q$ should be positive that means the energy transfer from dark energy to dark matter occurs. Thus, the positivity of the interaction term ensures that the second law of thermodynamics is realized \cite{pavon-2009}. In that case, since the unit $Q$ is the inverse of the cosmic time, it is natural for this value to be chosen as the product of the Hubble parameter and the energy density. Herein we consider $Q = 3 b^2 H \rho_{dm}$ in which $b^2$ is the coupling parameter or transfer strength. By inserting Eqs. \eqref{roheff1} into Eqs.\eqref{fried1} we obtain the energy density and the pressure of dark energy in the form
\begin{subequations}\label{fried2}
\begin{align}
 \kappa^2 \rho_{de}&= 3 H^2 \partial_R f -\frac{1}{2} R \partial_R f - \frac{1}{2} G \partial_G f + \frac{1}{2} f + 3 H \partial_R \dot{f} + 12 H^3 \partial_G \dot{f} - \kappa^2 \rho_{dm},\label{fried2-1}\\
\kappa^2 p_{de}&= \left(3 H^2 + \dot{H}\right) \partial_R f - \partial_R \ddot{f} - 2 H \partial_R \dot{f} - \frac{1}{2} f - 8 H \left(H^2 + \dot{H}\right) \partial_G \dot{f} - 4 H^2 \partial_G \ddot{f} \notag \\
&+ \frac{1}{2} G \partial_G f  - \kappa^2 (p_{dm} - 3 \xi H).\label{fried2-2}
\end{align}
\end{subequations}

The equation of state (EoS) of dark energy, $\omega_{de}$, is introduced as
\begin{equation}\label{eosde1}
\omega_{de} = \frac{p_{de}}{\rho_{de}},
\end{equation}
where the corresponding EoS of dark energy describes the late time dynamic properties of the universe which dependents on Gauss-Bonnet Gravity, dark matter, and viscous fluid. Therefore, in the next section, we will explore the dark matter EBEC.

\section{Dark matter as a EBEC}\label{III}

In this section, we are going to implement the concept of BEC for describing the nature of dark matter in the universe. As mentioned before, BEC is a kind of state of matter in which a dilute Bose gas is cooled into very low temperature that in this point phase transition and macroscopic quantum phenomenon happen as minimum quantum state. This evident make an effective connection between BEC and cosmic evolution of dark matter. It means dark matter is supposed to be a bosonic gas that constitutes BEC. Hence, one of the best models for describing dark matter can be BEC model. Now we introduce two approaches and explain them eloquently.

As the first approach, we consider the normal dark matter as bosonic particles in the early universe that comes from the concept of BEC. Therefore, the number density of particles follows from the Bose-Einstein statistics, which these particles are formed by decoupling of the remaining plasma in the early universe. Nevertheless, the energy density of dark matter is equal to the mass of dark matter multiplied by the number density of particles, and dark matter pressure is defined by Bose-Einstein statistics in a sphere with a radius of particles momentum (see Refs. \cite{Hogan-2000, Madsen-2001, Harko-2011, Craciun-2020} for more details). Hence, we can obtain the dark matter pressure in terms of the energy density of dark matter as a linear relationship in the following form
\begin{equation}\label{pressure1}
p_{dm}= \alpha \, \rho_{dm},
\end{equation}
where $\alpha$ \footnote[1]{$\alpha$ is written from two perspectives of normal dark matter and the kinetic theory of gases as $\alpha = <v^2>/3 = k_B T / m_{dm}$ \cite{Hogan-2000, Harko-2011, Chavanis-2017}.} is introduced as a multiple of the average squared velocity of the particles. Note that the above relation is inferred from the kinetic theory of gases, which is derived from the ideal law of gases in which no interaction between particles is considered. But from a cosmological point of view, the above equation leads to the barotropic EoS for the normal dark matter, which replaces with the dust-like theory of dark matter, i.e, $p_{dm} = 0$. Therefore, the barotropic EoS of the dark matter is proportional to the average squared velocity of the particles. Since dark matter is non-relativistic and follows from Bose-Einstein statistics, hence the variation of the numerical value of  the barotropic EoS has very little influence on the cosmic evolution of the universe. Therefore, Eq. \eqref{pressure1} acts as a non-interacting non-relativistic gas.

In the second approach, we consider the BEC dark matter as non-relativistic bosons with a two-particles interaction in a quantum system. As we know, BEC is a kind of state of matter in which a dilute Bose gas is cooled to very low temperatures. At this point, phase transition happens and a large part of bosons occupies a minimum quantum state, which means one condenses to the same quantum ground state. Thus, at absolute zero temperature, the dark matter is supposed as BEC matter halos in the hydrodynamic representation. In that case, the physical properties of the BEC are described by the generalized Gross-Pitaevskii equation \cite{Pitaevskii_2003, Pethick_2008}. Nevertheless, by considering BEC in the gravitational form, the corresponding EoS for BEC dark matter is written in the following form
\begin{equation}\label{pressure2}
p_{dm} = \beta  \rho_{dm}^2,
\end{equation}
where $\beta$ \footnote[2]{$\beta = 2 \pi \hbar^2 l_s / m_{dm}^3$ in which $l_s$ is the scattering length.} is introduced as a coefficient that relate to scattering length and mass of dark matter \cite{Harko-2011, Chavanis-2017, Harko1-2015}.

Now in order to have a deeper understanding of the universe, we extend the nature of dark matter from the perspective of BEC with the above approaches. In that case, for describing entire the epoch of the universe i.e., from early time till late time, we consider an extended form entitled EBEC for dark matter EoS as
\begin{equation}\label{pressure3}
p_{dm} = \alpha \, \rho_{dm} + \beta \, \rho_{dm}^2,
\end{equation}
where $\alpha$ and $\beta$ are respectively introduced as single-body interaction and two-body interaction, resulting from normal dark matter and dark matter halo, respectively. In what follows, we can see that $\alpha$ and $\beta$ play an important role in the nature of dark matter inside the universe. Hence, we will have the following cases for coefficients of EBEC dark matter:

1) If $\alpha$ and $\beta$ become zero, the dark matter pressure equal to zero, which indicates a cold dark matter case.

2) If $\beta$ becomes zero, we will have normal dark matter.

3) If $\alpha =0$, we have a hydrodynamic system as dark matter halo.

4) If $\alpha$ and $\beta$ become non-zero, we have the contribution of both cases the normal dark matter and dark matter halo.

On the other hand, we can express Eq. \eqref{pressure3} from the perspective of a particular EoS of the virial expansion. For this purpose, the virial expansion helps us to write the dark matter pressure as a power series in terms of the energy density of dark matter in the following form
\begin{equation}\label{pressure4}
\frac{p_{dm}}{  \alpha \rho_{dm}} = 1 + B \rho_{dm} + C \rho_{dm}^2 + D \rho_{dm}^3 + \cdots,
\end{equation}
where the first term represents normal dark matter, the second term represents the dark matter halo in a quantum ground state, and the third term and above are related to the excited quantum system of the dark matter halo. Therefore, in this paper, we only consider the first two terms of the virial expansion that including both single-body interaction and two-body interaction, then by this formalism, we take $\alpha = \alpha$, $B = \beta / \alpha$, and $C = D = \cdots = 0$.

In order to obtain the energy density of dark matter, we substitute Eq. \eqref{pressure3} into \eqref{contineq2-1} and will have
\begin{equation}\label{rhoch1}
\rho_{dm} = \rho_0\, \frac{c\, \eta \, a_0^{3 \eta} - \beta}{c\, \eta \, a^{3 \eta} - \beta},
\end{equation}
where $\rho_0 = \eta /(c\, \eta\, a_0^{3 \eta} - \beta)$, $a_0$, and $c$ are respectively the present energy density of dark matter, the current scale factor, and an integral constant, in which $\eta = \alpha + 1 - b^2$. It should be noted that this relationship shows that it depends on the scale factor, interacting term, and coefficients of EBEC dark matter.

\section{Reconstruction and Hubble data constraints}\label{IV}

In this section, we intend to investigate accelerated expansion of the universe by $f(R,G)$ gravity in an EBEC dark matter. This study helps us that encounter a more realistic model of the universe, which including the dark parts of the universe. This means that dark energy arises from $f(R,G)$ gravity, and dark matter comes from EBEC, also will have a good correspondence to the observational Hubble data.

Now in order to solve the corresponding Friedmann equations by findings of the previous section, we present the following conventional calculation as
\begin{subequations}\label{relations1}
\begin{align}
\frac{d f}{dt} &= \dot{R} \, \partial_R f + \dot{G} \, \partial_G f,\label{relations1-1}\\
\dot{R}& = 24 H \dot{H} + 6 \ddot{H},\label{relations1-2}\\
\dot{G} &= 96 H^3 \dot{H} + 24 H^2 \ddot{H} + 48 H \dot{H}^2,\label{relations1-3}\\
\partial_X \dot{f}& = \dot{X} \, \partial_{XX} f + \dot{Y} \, \partial_{XY} f,\label{relations1-3}\\
\partial_X \ddot{f} &= \dot{X}^2 \, \partial_{XXX} f + 2 \dot{X} \dot{Y} \, \partial_{XXY} f + \dot{Y}^2 \, \partial_{XYY} f,\label{relations1-3}
\end{align}
\end{subequations}
where each index $X$ and $Y$ take the values $R$ and $G$. For this purpose, we rewrite Eqs. \eqref{fried2} as
\begin{subequations}\label{fried3}
  \begin{align}
      \kappa^2 \rho_{de} & = -3 \left(H^2 + \dot{H}\right) \left(\partial_R f + 4 H^2 \partial_G f\right) + \frac{1}{2} f + 144 H^2 \left(4 H^2 \dot{H} + \dot{H}^2 + H \ddot{H}\right) \partial_{RG}f \displaybreak[0] \notag\\
        & + 18 H \left(4 H \dot{H} + \ddot{H}\right) \partial_{RR} f + 288 H^4 \left(4 H^2 \dot{H} + 2 \dot{H}^2 + H \ddot{H}\right) \partial_{GG} f - \kappa^2 \rho_{dm}, \label{fried3-1}\displaybreak[0]\\
      \kappa^2 p_{de} & = \left(3 H^2 + \dot{H}\right) \partial_R f + 12 H^2 \left(H^2 + \dot{H}\right) \partial_G f - \frac{1}{2} f \displaybreak[0] \notag\\
        &- 48 H  \left(8 H^3 \dot{H} + 6 H \dot{H}^2 + 2 H^2 \ddot{H} + \dot{H} \ddot{H}\right) \partial_{RG} f - 12 H \left(4 H \dot{H} + \ddot{H}\right) \partial_{RR} f \displaybreak[0] \notag\\
        &- 192 H^2 \left(H^2 + \dot{H}\right)\left(4 H^2 \dot{H} + H \ddot{H} + 2 \dot{H}^2\right) \partial_{GG} f - 36 \left(4 H \dot{H} + \ddot{H}\right)^2 \partial_{RRR} f  \displaybreak[0] \notag\\
        &- 144 H \left(4 H \dot{H} + \ddot{H}\right)\left(12 H^2 \dot{H} + 3 H \ddot{H} + 4 \dot{H}^2\right) \partial_{RRG} f \displaybreak[0] \notag\\
       &+ 2304 H^2 \dot{H}^2 \left(4 H^2 \dot{H} + H \ddot{H} + 2 \dot{H}^2\right) \partial_{RGG} f  \displaybreak[0] \notag\\
       &- 2304 H^4 \left(4 H^2 \dot{H} + H \ddot{H} + 2 \dot{H}^2\right)^2 \partial_{GGG} f - \kappa^2 (\alpha\,  \rho_{dm} + \beta\, \rho_{dm}^2 - 3 \xi H),\label{fried3-2}
  \end{align}
\end{subequations}
where these cosmological quantities of dark energy are related to the Hubble parameter and its higher derivatives, the form of function $f(R, G)$, EBEC dark matter, and viscous fluid.

In what follows, we reconstruct the above cosmological quantities with respect to redshift parameter $z$. In this case, by adopting the relation $a(t) = a_0 / (1+z)$ in which $a_0$ is the current scale factor, conversion relationships between cosmic time and redshift parameter are written as
\begin{subequations}\label{convert1}
\begin{align}
\frac{d}{dt} &= - H_0 (1+z) E(z) \frac{d}{dz}, \label{convert1-1}\\
\frac{d^2}{dt^2}& = H_0^2 (1+z)^2 E(z)^2 \frac{d^2}{dz^2} + H_0^2 (1+z)^2 E(z) E'(z) \frac{d}{dz} + H_0^2 (1+z) E(z)^2 \frac{d}{dz}, \label{convert1-2}
\end{align}
\end{subequations}
where the prime represents the derivative with respect to redshift parameter $z$, and also we take $H(z) = H_0 E(z)$ in which $H_0 = 67.4 \pm 0.5 \, km\,s^{-1}\,Mpc^{-1}$ is the present Hubble parameter \cite{Aghanim-2017}. Nevertheless, parameters $\dot{H}$ and $\ddot{H}$ are written in terms of $z$ in the form
\begin{subequations}\label{Hdot1}
\begin{align}
\dot{H} &= - H_0^2 (1+z) E E', \label{Hdot1-1}\\
\ddot{H} &= H_0^3 (1+z) E \left[(1+z) E E'' + (1+z) E'^2 + E E'\right]. \label{Hdot1-2}
\end{align}
\end{subequations}

By using the aforesaid relations, we reconstruct Eqs. \eqref{fried3} with respect to redshift parameter in the following form
\begin{subequations}\label{fried4}
\begin{align}
\kappa^2 \rho_{de}& = -3 H_0^2 E \left(1 - (1+z) E'\right) \left(\partial_R f + 4 H_0^2 E^2 \partial_G f\right) + \frac{1}{2} f \displaybreak[0] \notag \\
&+ 144 H_0^6 (1+z) E^4 \left(-3 E E' + 2 (1+z) E'^2 + (1+z) E E''\right) \partial_{RG}f \displaybreak[0] \notag \\
& + 18 H_0^4 (1+z) E^2 \left((1+z) E E'' + (1+z) E'^2 - 3 E E'\right) \partial_{RR} f \displaybreak[0] \notag \\
&+ 288 H_0^8 (1+z) E^6 \left((1+z) E E'' + 3 (1+z) E'^2 - 3 E E'\right) \partial_{GG} f - \kappa^2 \rho_{dm},\label{fried4-1}\\
\kappa^2 p_{de} &= H_0^2 E \left(3 E - (1+z) E'\right) \partial_R f + 12 H_0^4 E^3 \left(E - (1+z) E'\right) \partial_G f - \frac{1}{2} f \displaybreak[0] \notag \\
&+ 6 H_0^4 (1+z) E^2 \Bigl[(1+z) E E'' + (1+z) E'^2 - 3 E E'\Bigr]\partial_{RR} f \displaybreak[0] \notag \\
&+ 48 H_0^6 (1+z) E^3 \Biggl[6 E^2 E' - (1+z) E \Big( 7 E'^2 + 2 E E''\Big) + (1+z)^2 E' \Big(E E'' + E'^2\Big)\Biggr]\partial_{RG} f \displaybreak[0] \notag\\
&- 192 H_0^8 (1+z) E^5 \Big[E - (1+z) E'\Big] \Biggl[- 3 E E' + (1+z) E E'' + 3 (1+z) E'^2 \Biggr] \partial_{GG} f \displaybreak[0] \notag \\
&- 36 H_0^6 (1+z)^2 E^2 \Biggl[- 3 E E' + (1+z) E E'' + (1+z) E'^2 \Biggr]^2 \partial_{RRR} f \displaybreak[0] \notag \\
&- 144 H_0^8 (1+z)^2 E^4 \Big[3 E E' - (1+z) \left(E E'' + E'^2\right) \Big]\Biggl[9 E E' - (1+z) \Big(3 E E'' + 7 E'^2\Big) \Biggr] \partial_{RRG} f \displaybreak[0] \notag \\
&+ 2304 H_0^{10} (1+z)^3 E^6 E'^2 \Biggl[- 3 E E' + (1+z) E E'' + 3 (1+z) E'^2\Biggr] \partial_{RGG} f \displaybreak[0] \notag\\
&- 2304 H_0^8 (1+z) E^6 \Biggl[- 3 E E' + (1+z) E E'' + 3 (1+z) E'^2 \Biggr]^2 \partial_{GGG} f \displaybreak[0] \notag\\
&- \kappa^2 \left(\alpha\,  \rho_{dm} + \beta\, \rho_{dm}^2 - 3 \xi H_0 E\right).\label{fried4-2}
\end{align}
\end{subequations}

Now we consider a constraint for the Hubble parameter in terms of the redshift parameter by likelihood analysis with observational Hubble data. As we know, in general, one of the main solutions of the Friedmann equations is to write the first Friedmann equation in terms of density parameters, which we avoid in this study because of complexity of the equations in the our model. The advantage of this method is that the contribution of each component of the universe in terms of the density parameters and their fitting with the observational data can reach the most accurate answer. But because the study is in the late universe, we can ignore the contributions of radiation and neutrinos, which have negligible effect on the expansion of the universe. In the other hand, we consider the dark components of the universe as dark matter and dark energy that they come from EBEC model and $f(R, G)$ gravity, respectively. Therefore, the scenario of the power-law cosmology, which contains only one free parameter, is an appropriate alternative to solve the Friedmann equations which is written in the following form
\begin{equation}\label{at1}
a(t) = {a_0} \left(\frac{t}{t_0}\right)^m,
\end{equation}
where ${a_0}$ is the current scale factor, $t_0$ is the present age of the universe, and $m$ is a dimensionless positive coefficient \cite{Shafer-2015, Tutusaus-2016}. Then, we obviously obtain the Hubble parameter as
\begin{equation}\label{hubpar1}
H = \frac{m}{t},
\end{equation}
where we can find the current age of universe by inserting the present Hubble parameter $H_0$ into the above relation given by
\begin{equation}\label{hubpar2}
t_0 = \frac{m}{H_0},
\end{equation}
where coefficient $m$ is introduced as correction factor. So, we acquire the Hubble parameter in terms of the redshift parameter in the form
\begin{equation}\label{hubpar4}
H(z) = H_0 (1+z)^{\frac{1}{m}},
\end{equation}
where $E(z) = (1+z)^{\frac{1}{m}}$ is introduced as a dimensionless parameter for expansion rate.

\begin{figure}[t]
\begin{center}
\includegraphics[scale=.4]{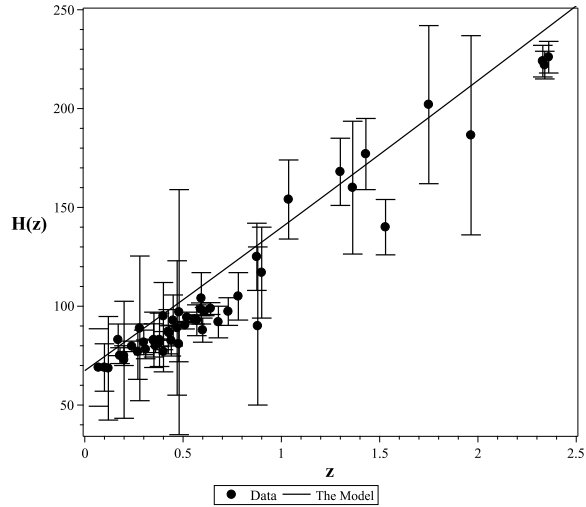}
\caption{The Hubble parameter in terms of the redshift data (circle + error bar) and the current model (line).}\label{fig1}
\end{center}
\end{figure}

Now we survey our model with 51 supernova data that the corresponding observational Hubble data came from the Refs. \cite{Blake_2012, Font_2014, Delubac_2015, Alam_2016, Moresco_2016, Farooq_2017, Pacif_2017, Magana_2018}. It should be noted that the Hubble parameter data is in the range of $0.07 \leq z \leq 2.36$, and measured by techniques of galaxy differential age or cosmic chronometer and radial BAO size methods. Therefore, we can use these data as the observational constraints for the attainment of the value $m$ in Eq. \eqref{hubpar4}. Note that since Eq. \eqref{hubpar4} has only one free parameter, it has less complexity and can be easily calculated with observational constraints. In what follows, the maximum likelihood analysis is introduced with the least-squares quantity so-called chi-square value ($\chi_{min}^2$) which is the simplest form of the observational data analysis. This is a very essential instrument for the fitting of data, especially in cosmology. In that case, the chi-square value $\chi_{min}^2$ for the Hubble data set is given by
\begin{equation} \label{likelihood1}
\chi_{min}^2 = \sum_{i = 1}^{51} \frac{\left(H_{obs}(z_i) - H_{th} (z_i, H_0)\right)^2}{\sigma_H^2 (z_i)},
\end{equation}
where $H_{obs}$ and $H_{th}$ indicate respectively the observed value and theoretical value of the Hubble parameter, $\sigma_H$ represents the standard error in the observed values. We can earn the value of correction factor as $m = 0.95$ from the best fitting with the Hubble parameter data set as shown in Fig. \ref{fig1}. We can see solid points as observational data and line graph as the best fit of our model in Fig. \ref{fig1}. Therefore, by substituting the obtained value of correction factor into Eq. \eqref{hubpar2}, the the age of universe obtains as $t_0 = 13.78 \, Gyr$. Thus, the mentioned astronomical data and test of measurement $t_0$ play an essential role in the observational constraints. Then Eq. \eqref{at1} can be a suitable choice for the study of the current job.

As we know, one of the methods for calculating the age of the universe is to find the values of the density parameters such as $\Omega_m$, $\Omega_r$, and so on, called baryonic matter, radiation, and so on. In addition to these parameters, the current Hubble parameter $H_{0}$ is the other important parameter for determining the age of the universe $t_0$ is given by
\begin{equation} \label{hubpar5}
t_0 = \frac{1}{H_0} F(\Omega_m, \Omega_r, \cdots),
\end{equation}
where $F$ is the function of the density parameters from various components within the universe. The first concept that we can understand from Eq. \eqref{hubpar5} is that the age of the universe is controlled by the Hubble parameter. In fact, the age of the universe value is a fraction of the inverse of the Hubble parameter, so in  the present job, we use more practical and achievable method. For this purpose, we take the correction factor $m$ instead of the function $F$ in the form of Eq. \eqref{hubpar2}. In that case if $m=1$, the age of the universe is $14.51 \, Gyr$ but this measurement does not correspond to the result obtained by observers. This means that the universe is younger for the Hubble parameter value as above mentioned namely $m = 0.95$.

\section{cosmological solutions and Stability analysis}\label{V}
In this section, we intend to solve the corresponding model by a particular regime. For this purpose, we consider a mixed power law model for viable $f(R,G)$ in the following form \cite{Dombriz-2012, Atazadeh-2014}
\begin{equation}\label{frg2}
f(R,G) = \lambda \, R^r \,  G^g,
\end{equation}
where $\lambda$, $r$ and $g$ are constants. Note that obtaining the form $f(R, G)$ is a problematic. One of the methods for obtaining the accurate form for function $f(R, G)$ is to consider the form of the scale factor function by an ansatz as Eq. \eqref{at1}. But due to the complexity of the Friedmann equations, it is practically tedious to obtain function $f(R, G)$. To this end, by mimicking cosmological solutions, we have considered the various forms for function $f(R, G)$ such as the additive law ($f(R, G) = r(R) + g(G)$) and the mixed power law (Eq. \eqref{frg2}), that choosing Eq. \eqref{frg2} can be the best option to continue. Nonetheless, $R$ and $G$ in Eqs. \eqref{RG1} are reconstructed in terms of redshift parameter in the following form
\begin{subequations}\label{convert1}
\begin{align}
R &= \frac{6 \,(2\, m - 1)}{m} H_0^2 E^2,\label{convert1-1}\\
G &= \frac{24\, (m-1)}{m^3} H_0^4 E^4.\label{convert1-2}
\end{align}
\end{subequations}

To substitute Eqs. \eqref{frg2} and \eqref{convert1} into Eqs. \eqref{fried4}, we can earn the energy density and pressure of dark energy as
\begin{subequations}\label{fried5}
\begin{align}
\kappa^2 \rho_{de}& = -3 H_0^2 E \left(1 - (1+z) E'\right) \left(\partial_R f + 4 H_0^2 E^2 \partial_G f\right) + \frac{1}{2} f \displaybreak[0] \notag \\
&+ 144 H_0^6 (1+z) E^4 \left(-3 E E' + 2 (1+z) E'^2 + (1+z) E E''\right) \partial_{RG}f \displaybreak[0] \notag \\
& + 18 H_0^4 (1+z) E^2 \left((1+z) E E'' + (1+z) E'^2 - 3 E E'\right) \partial_{RR} f \displaybreak[0] \notag \\
&+ 288 H_0^8 (1+z) E^6 \left((1+z) E E'' + 3 (1+z) E'^2 - 3 E E'\right) \partial_{GG} f - \kappa^2 \rho_{dm},\label{fried5-1}\displaybreak[0] \\
\kappa^2 p_{de}& = H_0^2 E \left(3 E - (1+z) E'\right) \partial_R f + 12 H_0^4 E^3 \left(E - (1+z) E'\right) \partial_G f - \frac{1}{2} f \displaybreak[0] \notag \\
&+ 6 H_0^4 (1+z) E^2 \Big[(1+z) E E'' + (1+z) E'^2 - 3 E E'\Big]\partial_{RR} f \displaybreak[0] \notag \\
&+ 48 H_0^6 (1+z) E^3 \Big[6 E^2 E' - (1+z) E \Big( 7 E'^2 + 2 E E''\Big) + (1+z)^2 E' \Big(E E'' + E'^2\Big)\Big]\partial_{RG} f \displaybreak[0] \notag \\
&- 192 H_0^8 (1+z) E^5 \Big[E - (1+z) E'\Big] \Big[- 3 E E' + (1+z) E E'' + 3 (1+z) E'^2 \Big] \partial_{GG} f \displaybreak[0] \notag \\
&- 36 H_0^6 (1+z)^2 E^2 \Big[- 3 E E' + (1+z) E E'' + (1+z) E'^2 \Big]^2 \partial_{RRR} f \displaybreak[0] \notag \\
&- 144 H_0^8 (1+z)^2 E^4 \Big[3 E E' - (1+z) \Big(E E'' + E'^2\Big) \Big]\Big[9 E E' - (1+z) \Big(3 E E'' + 7 E'^2\Big) \Big] \partial_{RRG} f \displaybreak[0] \notag \\
&+ 2304 H_0^{10} (1+z)^3 E^6 E'^2 \Big[- 3 E E' + (1+z) E E'' + 3 (1+z) E'^2\Big] \partial_{RGG} f \displaybreak[0] \notag \\
&- 2304 H_0^8 (1+z) E^6 \Big[- 3 E E' + (1+z) E E'' + 3 (1+z) E'^2 \Big]^2 \partial_{GGG} f \displaybreak[0] \notag\\
&- \kappa^2 \left(\alpha\,  \rho_{dm} + \beta\, \rho_{dm}^2 - 3 \xi H_0 E\right).\label{fried5-2}
\end{align}
\end{subequations}

\begin{figure}[h]
\begin{center}
\includegraphics[scale=.35]{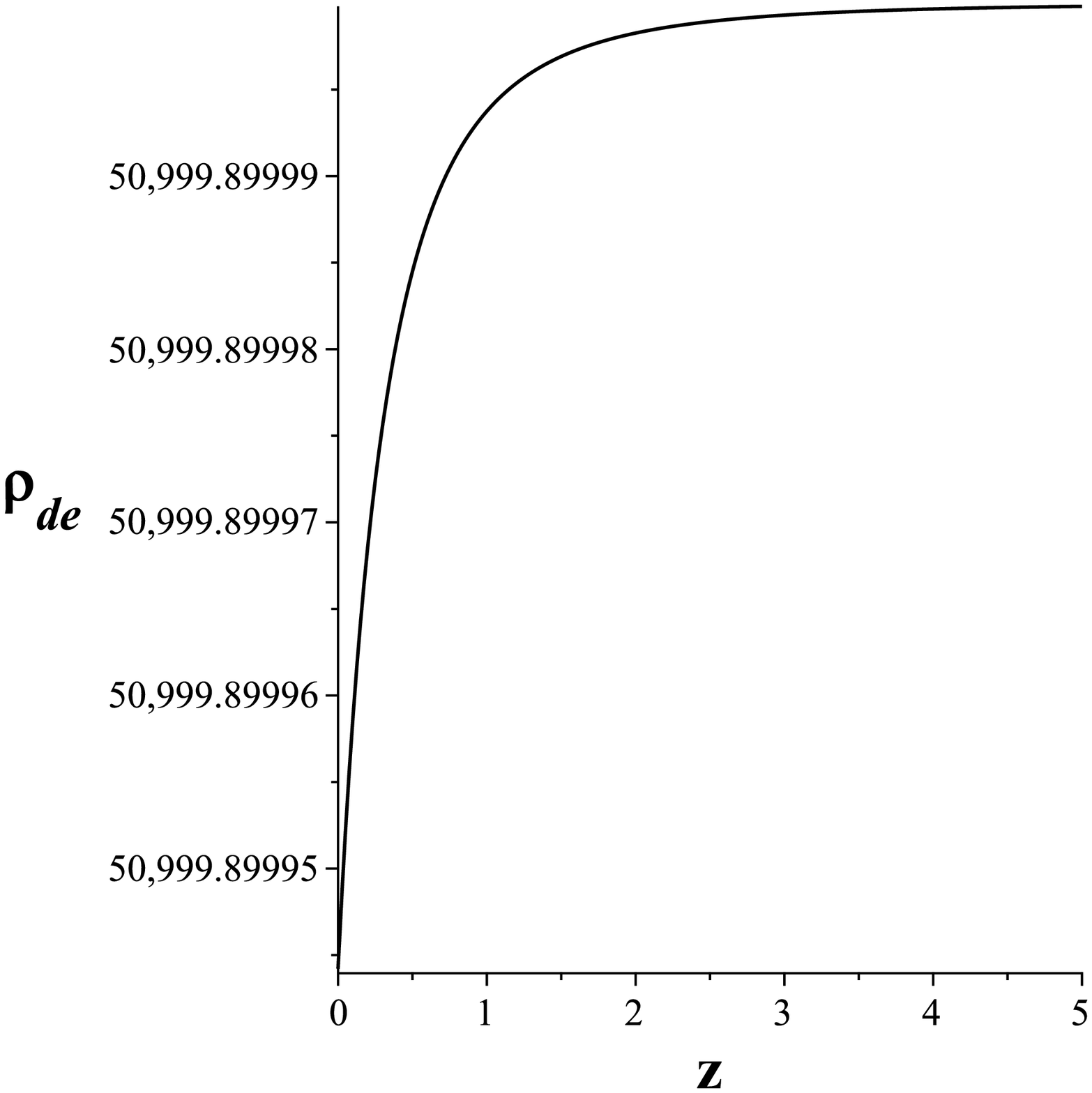}\quad\includegraphics[scale=.35]{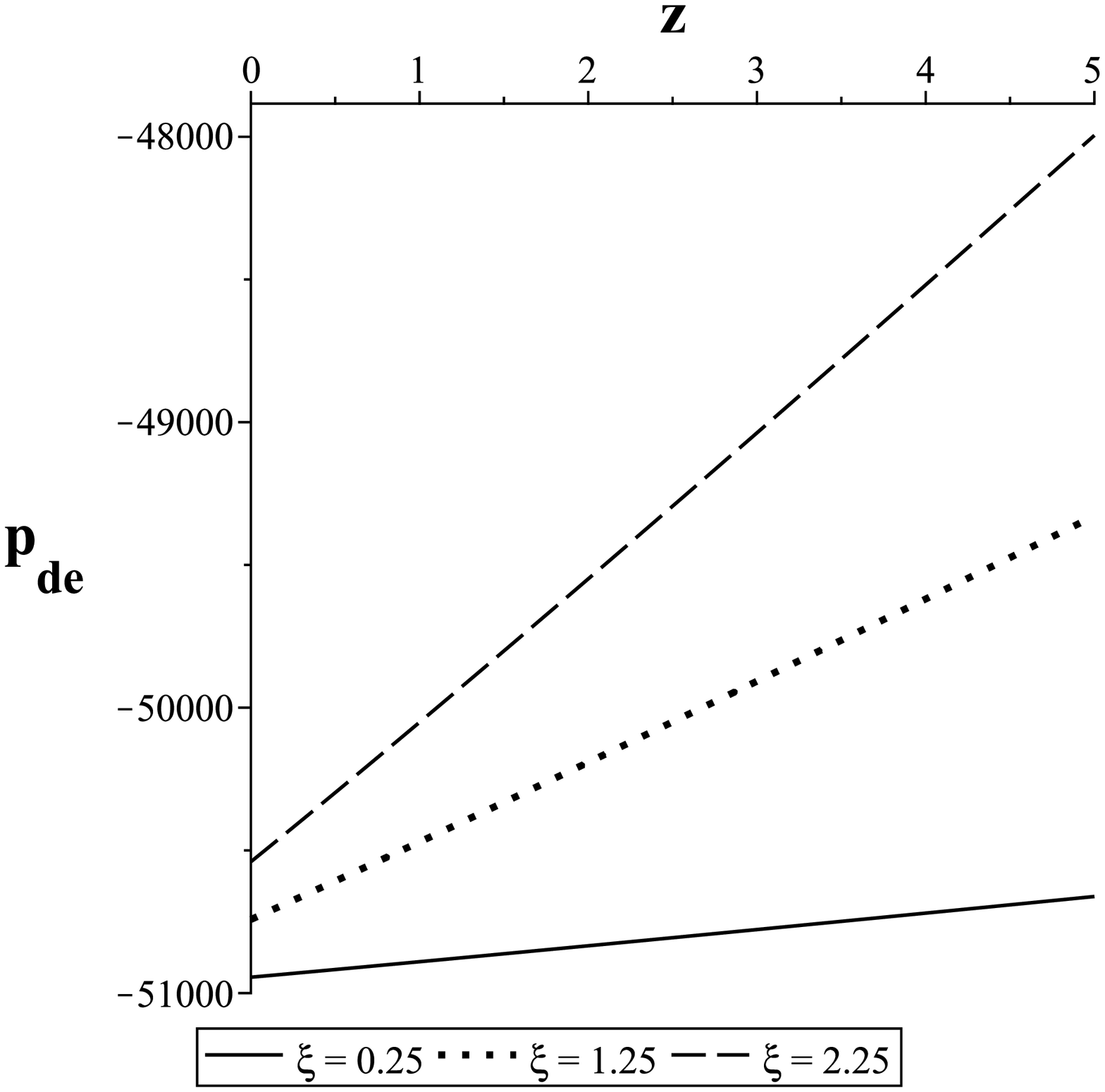}
\caption{The graphs of the energy density ($M L^{-1} T^{-2}$) and the pressure ($M L^{-1} T^{-2}$) of dark energy in terms of redshift parameter by $\alpha = 50$, $\beta = 0.001$, $b = 0.01$, $c = 2$, $m = 0.95$, $r = 0.5$, $g = -1$, $\lambda = 0.5$, $a_0 = 0.1$, and $\xi = 0.25, 1.25, 2.25$.}\label{fig2}
\end{center}
\end{figure}

\begin{figure}[h]
\begin{center}
\includegraphics[scale=.35]{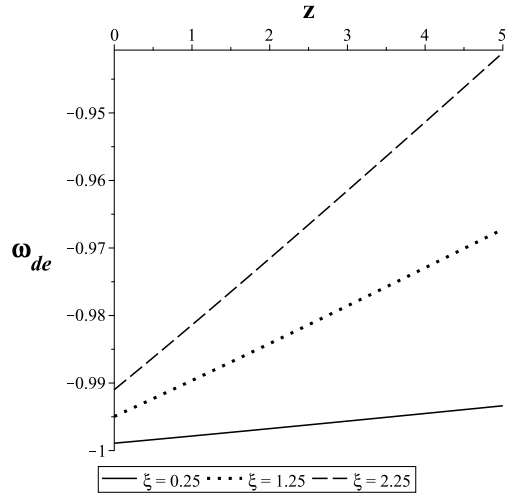}
\caption{The graph of the EoS (dimensionless quantity) of dark energy in terms of redshift parameter by $\alpha = 50$, $\beta = 0.001$, $b = 0.01$, $c = 2$, $m = 0.95$, $r = 0.5$, $g = -1$, $\lambda = 0.5$, $a_0 = 0.1$, and $\xi = 0.25, 1.25, 2.25$.}\label{fig3}
\end{center}
\end{figure}

By inserting corresponding equations such as $E$, $f$, and $\rho_{dm}$ into the aforesaid equation, the variety of energy density and pressure of dark energy are plotted in terms of redshift parameter as showed in Figs. \ref{fig2}. We must note that the role of free parameters is very important for description of dynamics of the cosmological parameters. Hence, their choice is very sensitive, i.e., there are different choices for the free parameters. For this purpose, our choices follow from the constraints such as $\rho_{de} > 0$ and $p_{de} < 0$ for the late universe ($z = 0$). Therefore, the free parameters in Figs. \ref{fig2} are drawn by $\alpha = 50$, $\beta = 0.001$, $b = 0.01$, $c = 2$, $m = 0.95$, $r = 0.5$, $g = -1$, $\lambda = 0.5$, $a_0 = 0.1$, and $\xi = 0.25, 1.25, 2.25$ which all are dimensionless quantities, except, parameter $\beta$ is dimension $M^{-4}$ in Planck units. Note that the units of the energy density and the pressure are both the same as $M L^{-1} T^{-2}$ in $SI$ metric and $M^4$ in Planck units. We can see in Figs. \ref{fig2} that the variety of the energy density depends to free parameters that come from $f(R, G)$ gravity, EBEC dark matter, and interacting model, as well as the variety of the pressure depends to $f(R, G)$ gravity, EBEC dark matter, interacting model, and viscous fluid. This figure shows us that the amounts of energy density decrease from a very high positive value in the early universe to a much lower positive value in the late universe ($z = 0$), also the amounts of pressure decrease from a very large positive value in the early universe to a much lower negative value in the late universe ($z = 0$).
Now we can draw EoS of dark energy \eqref{eosde1} in terms of redshift parameter by substituting Eqs. \eqref{fried5} into it which shown in Fig. \ref{fig3}. Note that the EoS parameter is a dimensionless quantity. In the modern cosmology, the EOS parameter is a very important parameter because it can describe various eras of the universe from the early time to the late time. In that case, $0 < \omega < \frac{1}{3}$ and $-1 < \omega < -\frac{1}{3}$ represent matter era and accelerated phase, respectively. Therefore, Fig. \ref{fig3} shows us that the universe begins from matter era to accelerated phase era in which the value of EoS is less than $-0.99$ for different values of viscosity in the present time ($z=0$), that represent the universe is undergoing an accelerated expansion. As is evident, the obtained value of dark energy EoS is confirmed by Refs. \cite{Amanullah_2010, Scolnic_2018}.

Now for a more complete evaluation, we explore the stability and the instability  of our model by using the sound speed parameter from the thermodynamics point of view. For this purpose, we consider that the universe is an adiabatic thermodynamics system, i.e., heat or mass doesn't transfer from inside the universe to its outside, so the entropy perturbation is equal to zero. In that case, we will only have the variation of the pressure in terms of the energy density, which leads to introduce the sound speed parameter $c_s^2$ in the following form
\begin{equation}\label{cs21}
c_s^2=\frac{\partial p_{de}}{\partial \rho_{de}} = \frac{\partial_z p_{de}}{\partial_z \rho_{de}},
\end{equation}
where $\partial_z = {\partial~}/{\partial z}$ represents derivative with respect to redshift parameter. Then, the corresponding parameter is a useful function to recognize the stability and instability. We note that conditions $c_s^2 > 0$ and $c_s^2 < 0$ represent the stability and instability, respectively.
\begin{figure}[h]
\begin{center}
\includegraphics[scale=.35]{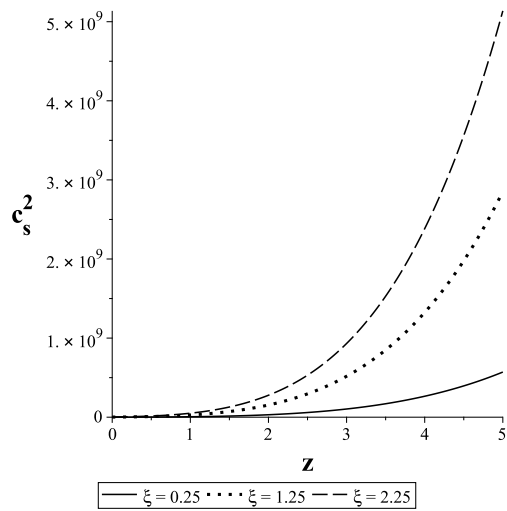}
\caption{The graph of the sound speed (dimensional quantity) in terms of redshift parameter by $\alpha = 50$, $\beta = 0.001$, $b = 0.01$, $c = 2$, $m = 0.95$, $r = 0.5$, $g = -1$, $\lambda = 0.5$, $a_0 = 0.1$, and $\xi = 0.25, 1.25, 2.25$.}\label{fig4}
\end{center}
\end{figure}
Now, we plot the variety of sound speed parameter in terms of redshift parameter as shown in Fig. \ref{fig4}. We can see from Fig. \ref{fig4} that generally the universe is in a stability phase from the early phase to the late time for the different values of viscosity. In the present model, the variation $c_s^2$ versus redshift parameter shows us that the energy density of dark energy leads to a controllable growing.

\section{Conclusion}\label{VI}
In this paper, we studied the dark matter BEC in $f(R, G)$ gravity by flat-FRW background in viscous fluid. Einstein equation has obtained by taking the variation from the action with respect to the metric, which represents the equivalent between the energy–momentum of matter and the geometry of space-time. Then, the effective energy density and the effective pressure have earned in terms of $f(R, G)$ gravity and viscous fluid. We considered that the content of the universe dominates the components of dark matter and dark energy which dark matter arises from the Extended Bose-Einstein density, and dark energy arises from $f(R, G)$ gravity and viscous fluid. After that the continuity equations have been written separately for the universe components by interacting between them. The interacting term is proportional with multiplied between the Hubble parameter and energy density of dark matter, i.e., there is the energy flow between dark matter and dark energy.

The another important subject in this paper is that EBEC can be an appropriate alternative for evolution of dark matter. BEC is a state of matter in which a dilute Bose gas cools to a very low temperature and reaches a ground quantum state. This regime suggests that dark matter is considered a bosonic gas. In what follows, we investigated that the EoS of dark matter based on BEC regime expressed different views such as $\omega_{dm} = p_{dm} / \rho_{dm} = \alpha$ and $\omega_{dm} = p_{dm} / \rho_{dm} = \beta \rho_{dm}$ where the first view is related to non-interacting bosonic particles, and the second view is related to two-particles interacting bosons in a quantum system. For this purpose, we considered a combination of the above two views as EBEC, this method helps us to earn more information about the nature of dark matter in entire the evolution of the universe. Nevertheless, the nature of one of the universe components, namely dark matter has been demonstrated by EBEC that lead to cognition of the late universe. On the other hand, EBEC model obeys the first and the second terms of the virial expansion that help us to write the EoS of dark matter, but the rest of the terms is concern to excited quantum system of the dark matter. The obtained results for $\alpha = 50$ and $\beta = 0.001$ represent that the quota of the single-body interaction is $50000$ times more than the quota of the two-body interaction.
Thus, the influence of non-interacting bosonic particles is more dominant than the influence of the bosonic particles with the quantum ground state.

After that, we reconstructed the corresponding relations in terms of the redshift parameter. This reconstruction represents that the cosmological parameters can be evaluated in the late time ($z = 0$). For this purpose, we considered the power-law cosmology for the scale factor, which obtained the Hubble parameter in terms of the redshift parameter. Then, we fitted the corresponding Hubble parameter relation by using the minimum chi-square value with 51 Hubble data constraints. Next, we plotted the figures of the cosmological parameters in terms of the redshift parameter by depending on the universe components. These figures showed us that the universe is undergoing accelerated expansion. At the end, we investigated stability analysis of the present model, that the obtained results indicated the universe is in a stability phase for the present time. This means that the growth of the energy density of dark energy is controlled in the late universe.


\end{document}